\documentclass[aps,%
  pra,%
  twocolumn,%
  groupedaddress,%
  showpacs,%
  amsmath,%
  amssymb,%
  ]{revtex4}

\usepackage{graphicx,color}
\usepackage{bm}

\begin{document}

\title{Paraxial Skyrmionic Beams}

\author{Sijia Gao}
\email{s.gao.2@research.glasgow.ac.uk} \affiliation{School of Physics and
Astronomy, University of Glasgow, Glasgow G12 8QQ, UK}
\author{J\"{o}rg B. G\"{o}tte}
\affiliation{School of Physics and
Astronomy, University of Glasgow, Glasgow G12 8QQ, UK}
\affiliation{College of Engineering and Applied Sciences,
Nanjing University,  Nanjing 210093, China}
\author{Fiona C. Speirits}
\author{Francesco Castellucci}
\author{Sonja Franke-Arnold}
\author{Stephen M. Barnett}
\email{Stephen.Barnett@glasgow.ac.uk} \affiliation{School of Physics and
Astronomy, University of Glasgow, Glasgow G12 8QQ, UK}

\date{\today}

\begin{abstract}
  Vector vortex beams possess a topological property that derives both from
  the spatially varying amplitude of the field and also from its varying
  polarization. This property arises as a consequence of the inherent
  Skyrmionic nature of such beams and is quantified by the associated
  Skyrmion number, which embodies a topological property of the beam. We
  illustrate this idea for some of the simplest vector beams and discuss the
  physical significance of the Skyrmion number in this context.
\end{abstract}


\maketitle


\section{Introduction}

Recent developments have highlighted the growing utility of structured light,
that is optical fields in which the spatial variation of the field amplitude
and/or the polarization are specifically designed for a given task
\cite{Nye,NyeBerry,Roberta,Dennis,Halina}.  Important examples include the
formation of optical beams carrying orbital angular momentum
\cite{Allen,AllenBook,Bekshaev,Sonja,Alison}, polarization or helicity patterns
\cite{CohenTPhillips,Dalibard,Rob12,Rob13,Koen,Kravets} and the vector vortex
beams and their relatives
\cite{Zhan,Piccirillo,Milione,Milione2,DAmbrosio,Bienvenu,Alpmann,Neal,Guzman}.
We show that there is a Skyrmion field associated specifically with vector
vortex beams and that the associated Skyrmion number is readily identified with
a simple property of the beam. As such the Skyrmion number provides a natural
way to present the variety of possible vector beams. It is noteworthy that
this property is explicitly a feature of vector beam: a Skyrmion
field exists only if both the polarization and the field amplitude are spatially
varying.

Skyrmions were first proposed for the study of mesons~\cite{Skyrme61,Skyrme62},
but the idea has since found wide application in many areas of physics
including quantum liquids~\cite{Vollhardt,Volovik,Leggett}, magnetic
materials~\cite{Sachdev,Seki,Dennis2}, 2D photonic materials~\cite{Mechelen}
and in the study of fractional statistics~\cite{Wilczek}. Recently they have
been observed in optics by the controlled interference of plasmon polaritons
\cite{Tsesses,Du}. We show here that a wide range of freely propagating optical
beams also possess a non-trivial Skyrmion field and with it a Skyrmion number,
the value of which is simply related to a topological property of the beam.


\section{Constructing Skyrmionic beams}

We consider a paraxial beam of either light~\cite{Marcuse,Siegman} or electrons
\cite{El-Kareh,Klemperer,Hawkes} and express the local polarization or spin
direction, respectively, in the form \begin{equation} \label{eq:state} |
\Psi(\bm{r})\rangle = u_0(\bm{r}) | 0 \rangle + e^{i \theta_0} u_1(\bm{r}) |1
\rangle \, .
\end{equation} Here $|0\rangle$ and $|1\rangle$ represent any two orthogonal
polarization (or electron spin) states, while $u_0(\bm{r})$ and $u_1(\bm{r})$
are two orthogonal spatial modes~\footnote{We have used a quantum mechanical
notation as this aids the analysis to come, but our results apply both to
classical and quantum states of light and also to electron beams.} and the
global phase difference between the two modes is denoted by $\theta_0$. That
this decomposition is always possible follows from the Schmidt
decomposition~\cite{QIbook}.  The Skyrmion field and number depend only on the
spatial variation of the polarization or spin direction and for this reason it
is convenient to work with a locally-normalized state in the form
\begin{equation} \label{eq:normstate} |\psi(\bm{r})\rangle = \frac{|0\rangle +
e^{i \theta_0} v(\bm{r})|1\rangle}{\sqrt{1 + |v(\bm{r})|^2}} \, , \end{equation}
where $v(\bm{r}) = u_1(\bm{r})/u_0(\bm{r})$.

\begin{figure}[htbp]
\centering
\includegraphics[width=0.95\linewidth]{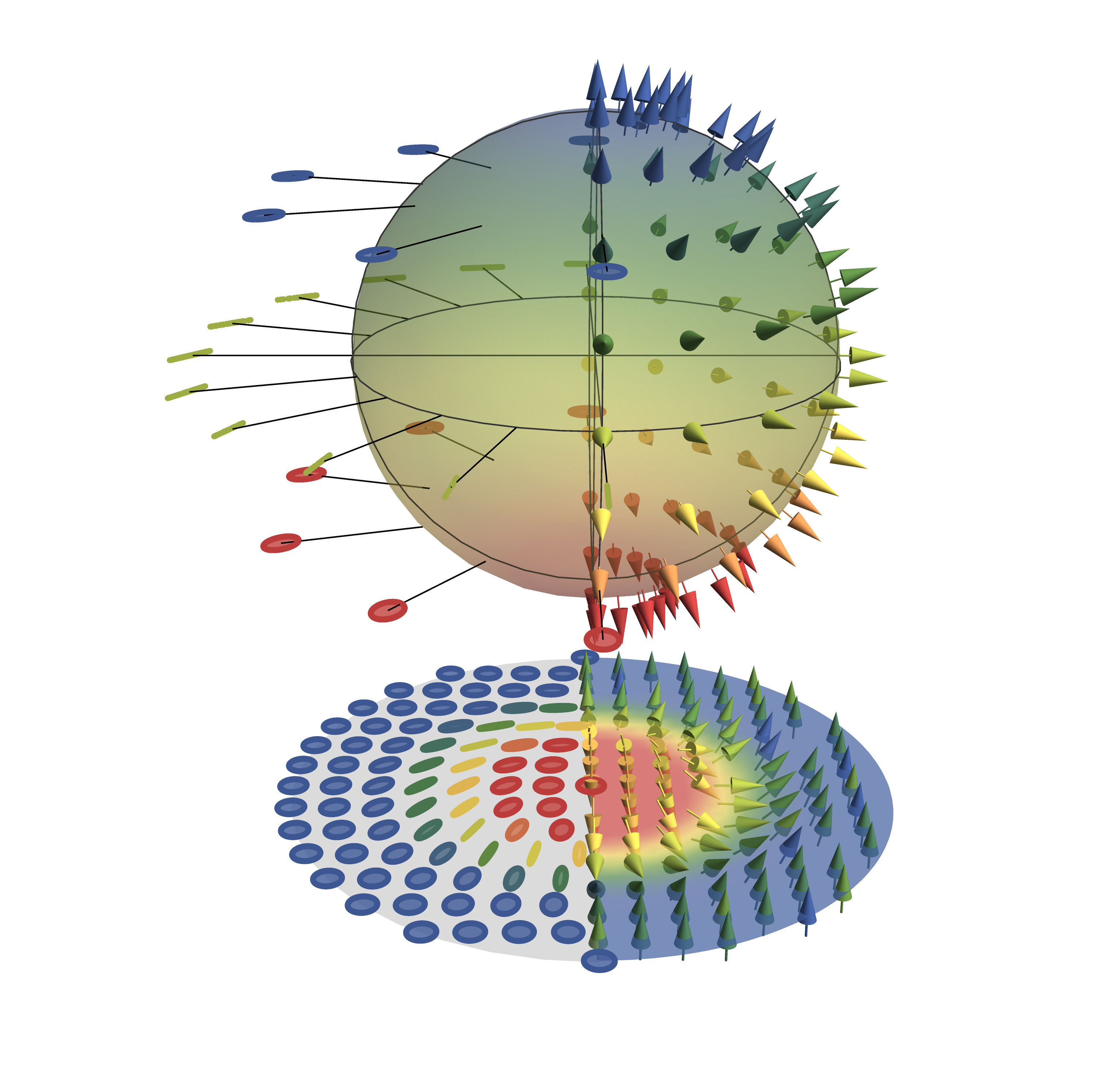}
\caption{Stereographic projection of the spatially varying polarization
or effective magnetization $\bm{M}$ onto the Poincar\'{e} or Bloch sphere.
We encode the degree of circular polarization $S_3$ and
the $z$ component $M_z$ on the same color scheme. For definiteness we choose in
our examples the polarization states $|0\rangle$ and $| 1 \rangle$ to correspond
to left and right handed circular polarization respectively or, for electrons,
the eigenstates of the $z$ component of the spin.} \label{fig:poincarebloch}
\end{figure}

The Skyrmion field is most readily defined in terms of an effective
magnetization $\bm{M}$, which is the local direction of the Poincar\'{e} vector
for light in Fig.~\ref{fig:poincarebloch} or the Bloch vector for an electron
beam. In terms of our locally normalized state it is
\begin{equation}
  \bm{M} = \langle\psi(\bm{r})|\bm{\sigma}|\psi(\bm{r})\rangle \, ,
\end{equation}
where $\bm{\sigma}$ is a vector operator with the Pauli matrices as
Cartesian components. For a light beam, the Cartesian components of
$\bm{M}$ correspond to the normalized local Stokes parameters $S_1, S_2$ and $S_3$
\cite{BornWolf}, and for the electrons to the local directions of the electron
spin.
The $i$th component of the associated Skyrmion field is
\begin{equation}
  \Sigma_i = \frac{1}{2}\epsilon_{ijk}\epsilon_{pqr}M_p\frac{\partial M_q}
  {\partial x_j}\frac{\partial M_r}{\partial x_k},
\end{equation}
where $\epsilon_{ijk}$ is the alternating or Levi-Civita symbol and we employ
the summation convention.
The form of the Skyrmion field ensures that it is transverse $(\bm{\nabla} \cdot
\bm{\Sigma} = 0)$. This means that there are no sources or sinks for the
Skyrmion field and the associated field lines can only form loops or extend to
infinity \footnote{This property is familiar from the study of electric and
magnetic fields $\bm{E}$ and $\bm{B}$ in free space. We note, however, that
$\bm{\Sigma}$ for light is a nonlinear function of $\bm{E}$ and $\bm{B}$ and so
its transverse nature is not simply a consequence of $\bm{\nabla} \cdot \bm{E} =
0 = \bm{\nabla} \cdot \bm{B}$.}.  It follows that the flux of the Skyrmion field
through any closed surface is zero, $\oint \bm{\Sigma} \cdot d\textbf{S} = 0$.

We consider a beam propagating in the $z$-direction. In each transverse plane of
the beam the polarization or spin pattern can form a Skyrmion reminiscent of
those familiar from the study of magnetic Skyrmions.
To facilitate this comparison, and also to characterize the variety of Skyrmions,
we employ the Skyrmion number
\begin{equation}
\label{eq:zcomponent}
 n = \frac{1}{4\pi}\int  \Sigma_z \: dxdy \, ,
\end{equation}
where the integral runs over the whole of the plane perpendicular to the
propagation direction of the beam.

Optical vector vortex beams typically have a spatially varying polarization pattern that
originates from the differential orbital angular momentum of the contributing
modes~\cite{Dennis,Piccirillo} and exhibit intriguing topological
\cite{Freund01,Dennis11,Freund11,Bauer} and focussing properties
\cite{Zhan02,Dorn}.  We consider the simplest case of such beams in which the
two orthogonal modes, with amplitudes $u_0(\bm{r})$ and $u_1(\bm{r})$, are
Laguerre-Gaussian (LG) modes
\begin{align}
  u^\ell_p (\rho,\phi,z) & =
  \sqrt{\frac{2p!}{\pi(p+|\ell|)!}}\frac{1}{w(z)} \left(\frac{\rho\sqrt{2}}{w(z)}
  \right)^{|\ell|} \exp\left(\frac{-\rho^2}{w^2(z)}\right) \nonumber \\
  & \times L^{|\ell|}_p\left(\frac{2\rho^2}{w^2(z)}\right) e^{i\ell\phi}
  \exp\left(-i \frac{\rho^2}{w^2(z)} \frac{z-z_0}{z_R} \right)  \nonumber \\
  & \times \exp\left[-i(2p+|\ell|+1)\tan^{-1}\left(\frac{z-z_0}{z_R}\right)
  \right] \, ,
\end{align}
familiar from the study of orbital angular momentum
\cite{Allen,AllenBook,Bekshaev,Sonja,Alison}. Here, we have employed cylindrical
polar coordinates ($\rho,\phi$,$z$), $z_R = \pi w_0^2/\lambda$ is the
Rayleigh range and $w(z) = w_0\sqrt{1 + (z-z_0)^2/z^2_R}$ is the beam width on
propagation. We assume that the modes have the same wavelength $\lambda$, but
they may differ in the beam parameters $\ell, p, w_0$ and the focal position
$z_0$. These modes have a vortex of strength $\ell$ on the $z$-axis, which is
associated with a $z$-component of the orbital angular momentum of $\ell\hbar$
per photon (or electron)~\cite{Allen,AllenBook,Bekshaev,Sonja,Alison}.  Modes
with different angular momentum numbers $\ell$ are orthogonal and if we choose
two such modes for our two complex amplitudes $u_0$ and $u_1$ in
(\ref{eq:state}) then the function $v(\bm{r})$ in (\ref{eq:normstate}) for the
locally normalized state $|\psi(\bm{r})\rangle$ has the general form
\begin{equation}
  \label{eq:model}
  v(\bm{r}) = f(\rho,z)e^{i\theta(\rho,z)}e^{i\Delta\ell\phi}
  \, , \end{equation}
where $\Delta\ell = \ell_1 - \ell_0$, $f$ and $\theta$ are real functions of the
coordinates $\rho$ and $z$, and $\theta$ incorporates all phase terms including
$\theta_0$, the phase difference between the modes.  It is then straightforward
to calculate the Skyrmion field and from this the Skyrmion number for our vector
vortex beam.  We find the simple result that for such beams the Skyrmion number
is
\begin{equation}
  \label{eq:skyrmionno}
  n = \Delta\ell\left(\frac{1}{1 +f^2(0,z)} - \frac{1}{1 + f^2(\infty,z)}
  \right) \, .
\end{equation}
The value of $n$ is determined solely by which of the two modes $u_0(\bm{r})$
and $u_1(\bm{r})$ dominates on the $z$-axis, the location of the vortex, and
which dominates as $\rho$ tends to infinity.

\begin{figure}[htbp]
\centering
\includegraphics[width=0.95\linewidth]{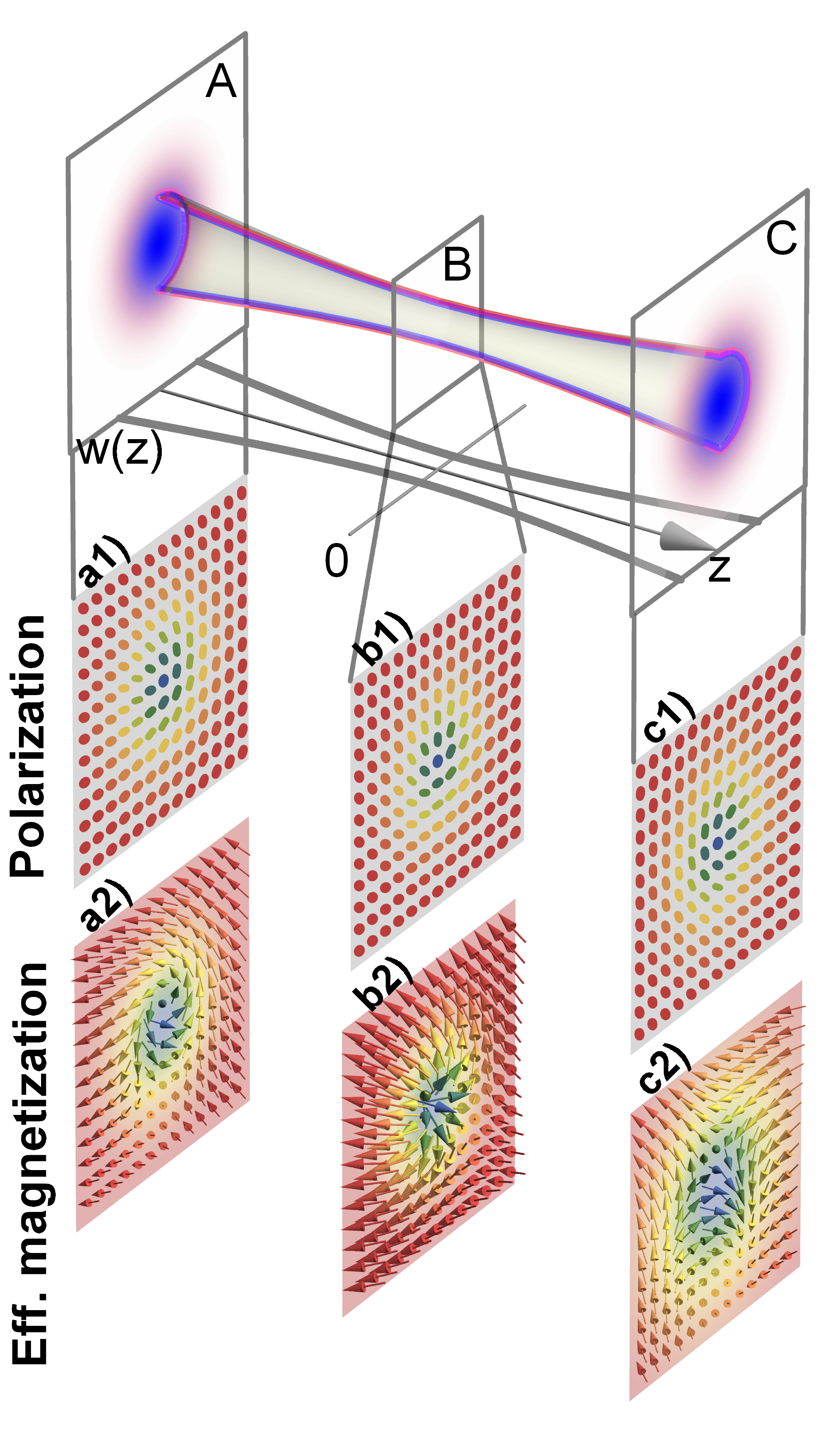}
\caption{Polarization structure for a superposition of LG modes with $\ell_1=1$
and  $\ell_0=0$ focussed at $z=0$.  The beam surface separating the regions in
which the modes have the larger amplitude, $u_0$ (blue) and $u_1$ (red). A, B, and C are three cross sections of interest, at $z=-10, 0$ and $10$ respectively. a1), b1) and
c1) are spatially varying polarization patterns coresponding to each plane while a2), b2) and c2) are the corresponding
effective magenitizations, with the classic chiral and hedgehog forms
respectively.} \label{fig:samefoc}
\end{figure}

The Skyrmionic beams that are simplest to construct comprise a superposition of
orthogonal polarization (or spin) states multiplied by $u^\ell_0$ LG modes with
no radial nodes, the
same beam width, a common focal point and with orbital angular momentum
differing by one.  In this case (\ref{eq:model}) simplifies to $v(\bm{r}) = A(z)
\rho e^{i\phi}$ (where $A$ is generally complex and includes the overall phase
difference $\theta_0$) and one polarization dominates at the position of the
vortex, with the orthogonal polarization appearing as $\rho\rightarrow\infty$.
We provide two examples of such polarization patterns in
Figs.~\ref{fig:samefoc}b \&~\ref{fig:samefoc}d together with the
corresponding effective magnetization in Figs.~\ref{fig:samefoc}c \&~\ref{fig:samefoc}e. The local Bloch vector, representing the
local spin direction, is
clearly reminiscent of the spiral and hedgehog Skyrmions, familiar from the
study of magnetic Skyrmions~\cite{Seki}; the former arises when $A$ is
imaginary and the latter when the amplitude $A$ is real.

We note that the natural propagation of the
beam will cause the magnetization or polarization pattern to evolve continuously from one of
these forms into the other by virtue of the relative Gouy phase~\cite{Siegman},
which changes as the beam propagates.
The Skyrmion number is unchanged, however, taking the value $+1$ at every
transverse plane.

There is however a subtle difference in the geometric interpretation between the
Poincar\'{e} and Bloch sphere. On both spheres, orthogonal states are
diametrically opposite. However, for the Poincar\'{e} sphere this corresponds to a
right angle in the major axes of the polarization, whereas the Bloch vectors
of orthogonal states are antiparallel.

\begin{figure}[htbp]
\centering \includegraphics[width=0.95\linewidth]{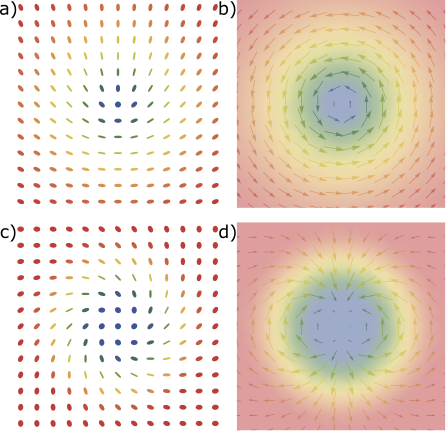}
\caption{Comparison of manifestations of spiral Skyrmions for polarization
and effective magnetization for two different Skyrmion numbers $n=1$ and $n=2$.
a) A spiralling polarization Skyrmion with $n=1$. The full rotation on
the Poincar\'{e} sphere results in half a rotation for the major axis of
the polarization ellipse.
b) The same configuration for the effective magnetization $\bm{M}$, where the
vector describes a full rotation on the Bloch sphere and in the configuration
space.
c) As in a) for $n=2$ showing now a full rotation of the polarization. d) As in
b) for $n=2$ showing two full rotations of the effective magnetization}
\label{fig:deltaell2}
\end{figure}

We can illustrate the effect of the discrepancy between rotation on the
Poincar\'{e} sphere and rotation of the polarization ellipse on the geometry
of the Skyrmion pattern in a comparision between spiral Skyrmions
from superposing LG beams with orbital angular momentum numbers
differing by one and two.  In Fig.~\ref{fig:deltaell2} we compare the local
polarization ellipse and Bloch vector for a pair of modes with $\Delta\ell = 1$
(as in Fig.~\ref{fig:samefoc}) with a pair of modes for which $\Delta\ell = 2$.  We see
that the polarization ellipses and the Bloch vectors rotate as one traverses a path
around the vortex. Moreover, along such a path, the polarization
ellipse completes half a rotation when $\Delta\ell = 1$, whereas the
Bloch vector rotates fully. For $\Delta\ell = 2$ the
polarization ellipse completes one full rotation and the Bloch vector winds
twice for one complete circle around the vortex.

These are examples of a more general result that for a
superposition of modes with a difference in orbital angular momentum number of
$\Delta\ell$, the Bloch or Poincar\'{e} vector rotates $\Delta\ell$ times on a
path enclosing the vortex. The corresponding polarization ellipse rotates by
only half the amount. This behavior persists when we consider modes with radial
indices different from zero, although the polarization structure becomes more
intricate because of the additional nodal lines. The resulting Skyrmion number
is nevertheless governed by the difference in dominating behavior described in
(\ref{eq:skyrmionno}).

\begin{figure}[t]
\centering
\includegraphics[width=0.95\linewidth]{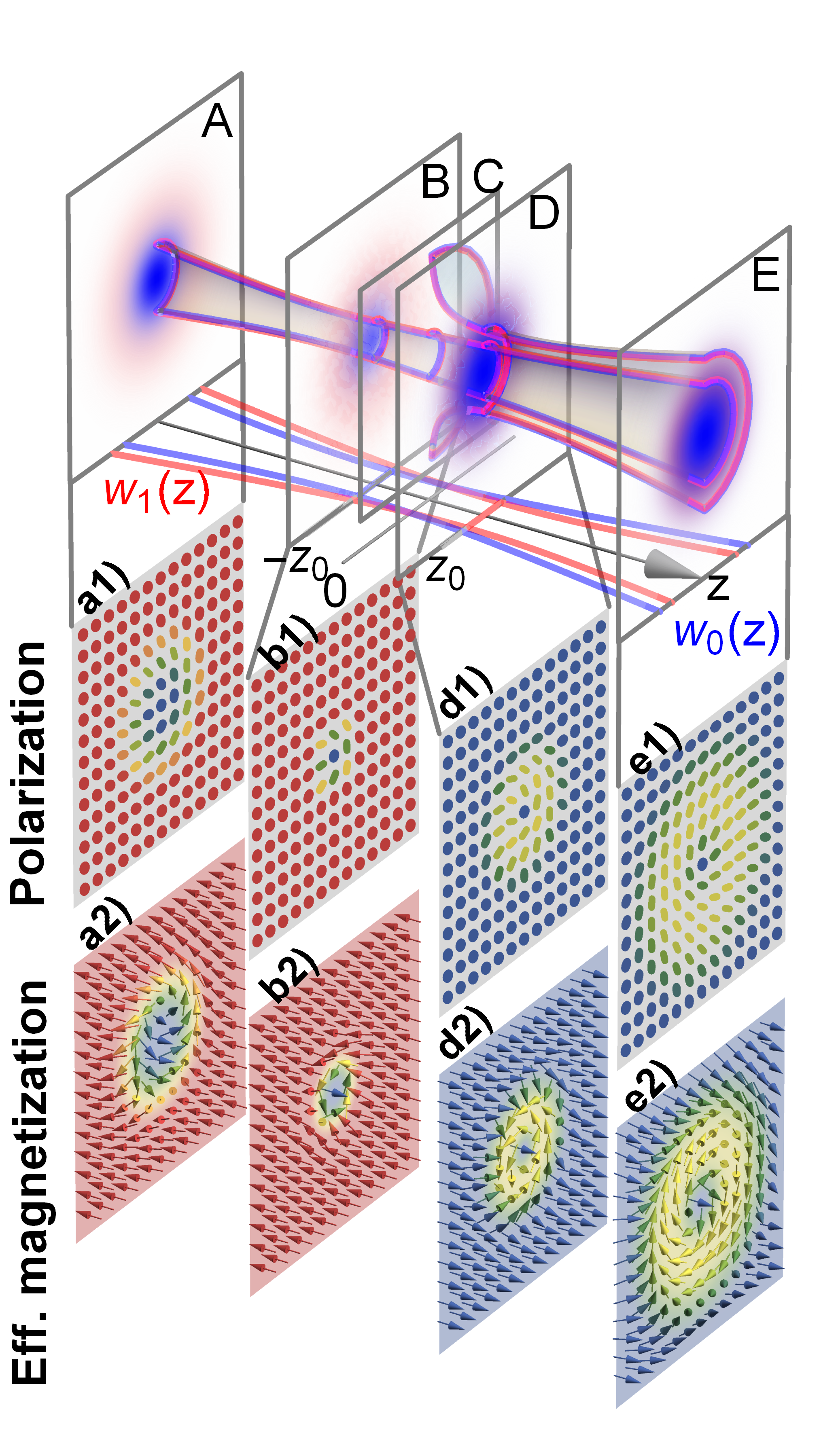}
\caption{Polarization structure for the same superposition of modes as in
Fig.~\ref{fig:samefoc} but focussed at different points $-z_0=-2$ and $z_0=2$.  For
$z>0$ the mode $u_0$ (blue) has the larger amplitude both in the central region
of the beam and the periphery as indicated by two surfaces of equal amplitude. A, B, D and E are four cross sections of interest, at $z=-10, -2, 2$ and $10$ respectively.
a1), b1) and d1), e1) are spatially varying polarization structure showing the qualitative
difference for $z \lessgtr 0$ at those four planes respectively. a2), b2) and d2), e2)  are the corresponding effective
magnetization with Skyrmion number 1 ($z < 0$) and 0 ($z > 0)$.}
\label{fig:diffoc}
\end{figure}

The corresponding Skyrmion number is $\Delta\ell$  if the spin or polarization
states at the vortex position and at infinity are orthogonal but will be zero if
they are the same. This dependence of the Skyrmion number on both $\Delta\ell$
and on the position dependence of the polarization clearly demonstrates that the
Skyrmion field and number are topological properties of both the spin and orbital
angular momenta.


\section{Conservation of the Skyrmion field}

The fact that the Skyrmion field, $\mbox{\boldmath$\Sigma$}$, is divergenceless
does not mean that the Skyrmion number, defined as the $z$-component of the flux
in (\ref{eq:zcomponent}) is necessarily conserved on propagation.
Consider a circular-cylindrical surface of radius $R$ centered on
the position of the vortex extending from $-z_0$ to $z_0$.
For the Skyrmion field to be divergenceless the flux through all surfaces of
this cylinder has to vanish.
The radial flux throug the mantle of the cylinder
\begin{align}
\label{eq:fluxrho}
\int_0^{2\pi} & d\phi \int_{-z_0}^{z_0} dz \, \Sigma_\rho = \nonumber \\
& -\Delta\ell\left(\frac{1}{1 + f^2(R,-z_0)} - \frac{1}{1 + f^2(R,z_0)}\right)
\end{align}
is compensated by the flux through the cylinder ends in the $z$-direction
at $z=-z_0$ and $z=z_0$. The expression for these is essentially given by
(\ref{eq:skyrmionno}), evaluated at $z=-z_0$ and $z=z_0$ and $\rho=R$
instead of infinity. The two terms evaluated $\rho=0$ for $z=-z_0$ and $z=z_0$
cancel and the total flux through both ends of the cylinder is given by
\begin{equation}
  \Delta\ell\left(\frac{1}{1 + f^2(R,-z_0)} - \frac{1}{1 + f^2(R,z_0)}\right),
\end{equation}
which is the the negative of (\ref{eq:fluxrho}), proving that there is no
total flux through the cylinder.

If we now construct a superposition of LG beams such that the
radial flux is non-vanishing, the flux along the $z$ direction also needs to
be different from zero which indicates a change in the Skyrmion number.
The simplest way to demonstrate this is to consider a superposition of LG beams
that are focussed at different positions along the $z$-axis.  The effect of
this is that the polarization behavior at large values of $\rho$ changes as
the beam propagates and the Skyrmion number changes from $\Delta\ell$ to
$0$ (or from $0$ to
$\Delta\ell$). This behavior is depicted in Fig.~\ref{fig:diffoc}, where we see
that the polarization at large distances from the central vortex changes
abruptly at one transverse plane and with it the Skyrmion number.
At plane A and B the Skyrmion number is $+1$ and at plane D and E it is equal to zero. The
boundary between these two regimes is at plane C, where the Skyrmion field
lines escape to $\rho \to \infty$.
Clearly, this will give a non-zero value for the radial flux because
$f^2(R,-z_0) \neq f^2(R,z_0)$ and hence a change in the Skyrmion number if
we allow $R$ to tend to infinity.


\section{Conclusions}

We have shown that paraxial vector vortex beams, either of light or electrons, possess a
topological property that can be identified with a Skyrmion number. The
associated Skyrmion field is transverse (or divergenceless) and this means that
there are no sinks or sources of this field. The Skyrmion number for a beam can
change on free space propagation, however, if Skyrmion field lines escape
radially out of the beam towards regions of negligible intensity. Demonstrating
these properties requires the preparation of vector vortex beams and measurement
of the polarization or spin in planes perpendicular to the beam axis
\cite{Sonja2}. We shall report on such experiments elsewhere.

We close by emphasising that the Skyrmionic property of vector beams is distinct
from the familiar spin and orbital angular momentum of optical beams
\cite{Allen,AllenBook,Bekshaev,Sonja,Alison,Barnett}. It is true that the beams
we consider here combine optical vortices and polarization, commonly associated
with orbital and spin angular momentum respectively, but the Skyrmion number is
a topological rather than a mechanical property of the beam. To see this we note
that the Skyrmion number is unchanged if we apply a global transformation of the
polarization, for example via reflection at a surface or a phase
retardiation of the constituent beams.  On the other hand we have seen that
it is possible for the Skyrmion number
to change if the two superimposed modes are focussed at different propagation
distances. The
total spin and angular momentum passing through each transverse plane, however,
remains unchanged.

\begin{acknowledgments}
This work was supported by the Royal Society (RP150122 and RPEA180010), and by
the UK Engineering and Physical Sciences Research Council (EP/R008264/1) and by the European Training Network ColOpt, funded by the European Union (EU) Horizon 2020 program under the Marie Sklodowska-Curie Action, Grant Agreement No.~721465.
\end{acknowledgments}

\bibliographystyle{apsrev}
\bibliography{skyrmions}

\end{document}